\documentclass[
preprint,
superscriptaddress,
 amsmath,amssymb,
 aps,
 prl,
 floatfix,
]{revtex4-2}

\usepackage{graphicx}% Include figure files
\usepackage{epstopdf} %converting to PDF
\usepackage{dcolumn}% Align table columns on decimal point
\usepackage{bm}% bold math
\usepackage{siunitx}
\usepackage{hyperref}

\begin{document}

%\preprint{APS/123-QED}
\title{Coupling of ferromagnetic and antiferromagnetic spin dynamics in Mn$_\mathrm{2}$Au/NiFe thin-film bilayers}% Force line breaks with \\
%\title{Magnetization dynamics}% Force line breaks with \\
\author{Hassan Al-Hamdo}
\email[E-Mail: ]{alhamdo@rhrk.uni-kl.de}
\affiliation{Fachbereich Physik and Landesforschungszentrum OPTIMAS, Rheinland-Pf{\"a}lzische Technische Universit{\"a}t Kaiserslautern-Landau, 67663 Kaiserslautern, Germany}

\author{Tobias Wagner}
\affiliation{Institute of Physics, Johannes Gutenberg-University Mainz, 55099 Mainz, Germany}

\author{Yaryna Lytvynenko}
\affiliation{Institute of Physics, Johannes Gutenberg-University Mainz, 55099 Mainz, Germany}
\affiliation{Institute of Magnetism of the NAS of Ukraine and MES of Ukraine, 03142 Kyiv, Ukraine}
\author{Gutenberg Kendzo}
\affiliation{Fachbereich Physik and Landesforschungszentrum OPTIMAS, Rheinland-Pf{\"a}lzische Technische Universit{\"a}t Kaiserslautern-Landau, 67663 Kaiserslautern, Germany}

\author{Sonka Reimers}
\affiliation{Institute of Physics, Johannes Gutenberg-University Mainz, 55099 Mainz, Germany}

\author{Moritz Ruhwedel}
\affiliation{Fachbereich Physik and Landesforschungszentrum OPTIMAS, Rheinland-Pf{\"a}lzische Technische Universit{\"a}t Kaiserslautern-Landau, 67663 Kaiserslautern, Germany}

\author{Misbah Yaqoob}
\affiliation{Fachbereich Physik and Landesforschungszentrum OPTIMAS, Rheinland-Pf{\"a}lzische Technische Universit{\"a}t Kaiserslautern-Landau, 67663 Kaiserslautern, Germany}

\author{Vitaliy I. Vasyuchka}
\affiliation{Fachbereich Physik and Landesforschungszentrum OPTIMAS, Rheinland-Pf{\"a}lzische Technische Universit{\"a}t Kaiserslautern-Landau, 67663 Kaiserslautern, Germany}

\author{Philipp Pirro}
\affiliation{Fachbereich Physik and Landesforschungszentrum OPTIMAS, Rheinland-Pf{\"a}lzische Technische Universit{\"a}t Kaiserslautern-Landau, 67663 Kaiserslautern, Germany}

\author{Jairo Sinova}
\affiliation{Institute of Physics, Johannes Gutenberg-University Mainz, 55099 Mainz, Germany}

\author{Mathias Kl{\"a}ui}
\affiliation{Institute of Physics, Johannes Gutenberg-University Mainz, 55099 Mainz, Germany}

\author{Martin Jourdan}
\affiliation{Institute of Physics, Johannes Gutenberg-University Mainz, 55099 Mainz, Germany}

\author{Olena Gomonay}
\affiliation{Institute of Physics, Johannes Gutenberg-University Mainz, 55099 Mainz, Germany}

\author{Mathias Weiler}
\affiliation{Fachbereich Physik and Landesforschungszentrum OPTIMAS, Rheinland-Pf{\"a}lzische Technische Universit{\"a}t Kaiserslautern-Landau, 67663 Kaiserslautern, Germany}

\begin{abstract}
We investigate magnetization dynamics of Mn$_\mathrm{2}$Au/Py (Ni$_\mathrm{80}$Fe$_\mathrm{20}$) thin film bilayers using broadband ferromagnetic resonance (FMR) and Brillouin light scattering spectroscopy. Our bilayers exhibit two resonant modes with zero-field frequencies up to almost \SI{40}{\giga\hertz}, far above the single-layer Py FMR. Our model calculations attribute these modes to the coupling of the Py FMR and the two antiferromagnetic resonance (AFMR) modes of Mn$_\mathrm{2}$Au. The coupling-strength is in the order of \SI{1.6}{\tesla \nano \meter} at room temperature for nm-thick Py. Our model reveals the dependence of the hybrid modes on the AFMR frequencies and interfacial coupling as well as the evanescent character of the spin waves that extend across the Mn$_\mathrm{2}$Au/Py interface. 

\end{abstract}

\maketitle

%\section{\label{Introduction}Introduction\protect}
Ferromagnets have a net magnetic moment and uniform spin dynamics in the GHz range~\cite{kittel_theory_1948}. In contrast, collinear  antiferromagnets have two equal but opposite sublattice magnetizations with vanishing net magnetic moment~\cite{neel_proprietes_1936, nagamiya_antiferromagnetism_1955, nagamiya_antiferromagnetism_1955} and spin dynamics that can reach the THz range~\cite{Rezende:Introduction:2019, kampfrath_coherent_2011}. In addition to THz spin dynamics, antiferromagnets can exhibit significant stability  of their magnetic moments against external magnetic perturbations~\cite{sapozhnik_direct_2018}. 
%Antiferromagnets can be controlled by spin currents, making them a promising alternative to heavy-metal-based spintronic devices~\cite{inbook,jungwirth2016antiferromagnetic,jungfleisch2018perspectives,RevModPhys.90.015005,PhysRevLett.113.157201,bodnar2018writing}.
The difference in the magnetization dynamics of ferromagnetic (FM) and antiferromagnetic (AFM) materials could potentially be exploited in applications that integrate AFM materials in high-frequency spintronic devices. A promising approach to enhance the FM spin dynamics frequencies and control FM spin-wave dispersions might be the combination of FM and AFM thin-film layers with interfacial exchange coupling. As a result of interfacial exchange coupling, a pronounced increase in the coercivity of the FM layer~\cite{noauthor_magnetization_2020,malozemoff_random-field_1987,stiles_model_1999,hoffmann_symmetry_2004, trassin_interfacial_2013} and exchange bias~\cite{meiklejohn_new_1956, NOGUES1999203, cai_exchange_nodate,lin_improved_1994,meiklejohn_new_1957,binek_training_2004,bhattacharya_antiferro-ferromagnetic_2017,Cord:Dynamic:2004} can be observed. Interfacial coupling also modifies GHz spin dynamics in FM/FM, chiral FM/FM and AFM/FM heterostructures, in particular the magnetic damping and anisotropy~\cite{Scott:Ferromagnetic:1985, weber_modified_2005, silsbee_coupling_1979, heinrich_dynamic_2003, Youssef:FMR:2016, BeikMohammadi:Broadband:2017, Camley:Probing:1999, Dantas:Ferromagnetic:2012, Hu:Ferromagnetic:2002, Layadi:Exchange:2002, Phuoc:Ultrahigh:2010, Qin:Exchangetorqueinduced:2018a, Polishchuk:Isotropic:2021, Inman:Hybrid:2022, Klingler:SpinTorque:2018, MacNeill:Gigahertz:2019b,  Li:Magnon:2020, Wang:Sub50:2021a, Wang:Hybridized:2022, Xiong:Tunable:2022, Luthi:Hybrid:2023}. However, the role of the THz frequency spin dynamics of AFMs for the hybrid spin dynamics in AFM/FM bilayers has not been revealed so far. Consequently, AFM/FM bilayers have so far not been leveraged to control hybrid mode frequencies or study AFM dynamics without requiring THz spectroscopy tools. 

Here we show that by coupling AFM and FM modes, we can make the AFM dynamics visible in the GHz range and use the exchange-enhancement~\cite{liensberger_exchange-enhanced_2019} of AFM modes to elevate the FM spin dynamics frequencies. The existence of such a coupling between FM and AFM spin dynamics opens new possibilities to design next generation magneto-electronic devices that exploit AFM materials beyond exchange bias. The Mn$_2$Au/Py system~\cite{ jourdan_epitaxial_2015} is particularly promising for studying the fundamental properties of interfacial coupling between FM and AFM thin films, as this system has a well-defined interface termination of AFM moments~\cite{bommanaboyena_readout_2021}. The coupling of the static Py magnetization to the Mn$_2$Au Néel vector can be used to control the AFM Néel vector orientation by magnetic fields even below \SI{1}{\tesla}~\cite{bommanaboyena_readout_2021, lytvynenko2022current}.

%The existence of interfacial exchange coupling can be used to engineer high-anisotropy materials. In such a system, the magnetic anisotropy energy at the interface is related to the coupling between the spin systems of the two layers, that results in a favored axial orientation of the spins~\cite{meiklejohn1957new}.
%The existence of such a coupling opens new possibilities to design new generation magneto-electronic devices because of its ability to change the magnetic properties of the FM layer~\cite{ jourdan2015epitaxial,meiklejohn1956new, nogues1999exchange, lin1994improved, bhattacharya2017antiferro, article,inbook}. The Mn$_2$Au/Py system is particularly promising for studying the fundamental properties of interfacial coupling between FM and AFM thin films, as this system has a well-defined interface termination of AFM moments. The coupling of the static Py magnetization to the Mn$_2$Au Néel vector can be used to control the AFM Néel vector orientation by magnetic fields much smaller than \SI{1}{\tesla}~\cite{bommanaboyena2021readout, lytvynenko2022current}.  

We investigate the resonant magnetic dynamics of a hybrid system consisting of thin-film poly-crystalline FM layers (Py) deposited on single-crystalline thin-film antiferromagnets (Mn$_\mathrm{2}$Au). We study the quasi-uniform dynamics by vector-network-analyzer ferromagnetic resonance (VNA-FMR) and we study the spin-wave response by Brillouin Light Scattering (BLS). We model the observed two eigenmodes of the hybrid system in the context of evanescent spin-wave modes that extend from the Py layer into the Mn$_2$Au layer and which are coupled to the two non-degenerate modes of the easy-plane antiferromagnet.  

\begin{figure}
\centering
\includegraphics{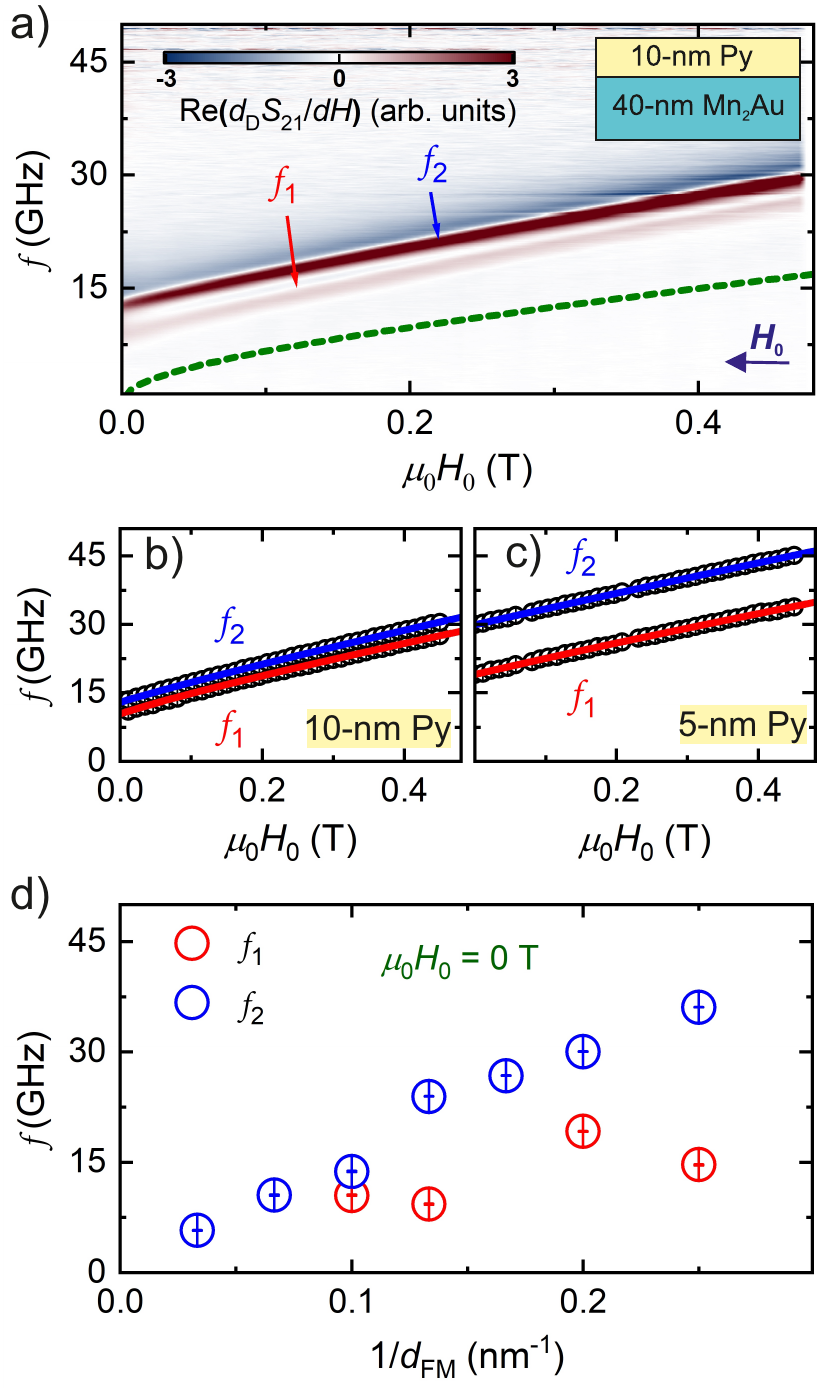}
\caption{(a) Real part of the background corrected VNA signal of a Mn$_\mathrm{2}$Au (40 nm)/Py (10 nm) sample. The dashed line indicates the expected FMR for a bare Py thin film. (b) Resonance frequencies $f_1$ and $f_2$  vs. external magnetic field $H_0$ obtained from fitting data in panel (a). The lines are fits to the modified Kittel equation from our model (see text).
%with M$_\mathrm{FM}= 1$ T as a fitting parameter. 
(c) Same as (b) but for a Mn$_\mathrm{2}$Au (40 nm)/Py (5 nm) sample. (d) The zero-field mode frequencies $f_1(H_0=0)$ and $f_2(H_0=0)$ increase with increasing $d_\mathrm{FM}^{-1}$.}~\label{fig:BBFMR1}
\end{figure}
To carry out this investigation, epitaxial Mn$_\mathrm{2}$Au  thin films with a thickness of 40 nm, and  Py with variable thicknesses $\SI{2}{\nano\meter}\leq d_\mathrm{FM} \leq \SI{30}{\nano\meter}$ are deposited on Al$_\mathrm{2}$O$_\mathrm{3}$ substrate~\cite{jourdan_epitaxial_2015,bommanaboyena_readout_2021,sapozhnik_experimental_2018}. To determine the resonance frequency of the uniform modes in the samples we performed VNA-FMR in the frequency sweep mode. The external magnetic field  is applied in the sample plane (for more details see \cite{SI}). %The $S_{21}$-parameter was measured as a function of frequency for each external magnetic field value .

Figure~\ref{fig:BBFMR1}(a) shows the background-corrected $\mathrm{Re}(d\textsubscript{D} S\textsubscript{21}/dH)$ VNA-FMR data~\cite{maier-flaig_note_2018} obtained for a Mn$_\mathrm{2}$Au (40 nm)/Py (10 nm) sample. We observe two distinct resonance modes,  a faint mode with frequency $f_1$ and a stronger mode with frequency $f_2>f_1$. The mode frequencies are enhanced by about 10 GHz compared to the uncoupled FMR frequency for in-plane isotropic polycrystalline Py (dashed line). In contrast to earlier studies that found an enhancement of the FMR frequency of a similar magnitude in FM/AFM bilayers~\cite{Phuoc:Ultrahigh:2010, Phuoc:Influence:2011}, we observe two distinct modes. We fit the obtained complex-valued $d\textsubscript{D} S\textsubscript{21}$ spectra as a function of frequency for a series of $H_0$-values by the sum of two magnetic resonances as described in the SI~\cite{SI}. From the fits we obtain the two resonance frequencies $f_1$ and $f_2$ as well as the corresponding linewidths and spectral weights. We show the fitted $f_1$ and $f_2$ in Fig.~\ref{fig:BBFMR1}(b). We perform identical measurements for a series of  Mn$_\mathrm{2}$Au(\SI{40}{\nano\meter})/Py($d_\mathrm{FM}$) samples, where the Py thickness $\SI{4}{\nano\meter}\leq d_\mathrm{FM} \leq\SI{30}{\nano\meter}$ is varied. 
%We show the extracted $f_1$ and $f_2$ for a sample with $d_\mathrm{FM}=\SI{5}{\nano\meter}$ in Fig.~\ref{fig:BBFMR1}(c) while the full corresponding dataset can be found in the SI \cite{SI}. 
When changing the Py thickness from $d_\mathrm{FM}=\SI{10}{\nano\meter}$ to $d_\mathrm{FM}=\SI{5}{\nano\meter}$, we observe that both, $f_1$ and $f_2$, are clearly enhanced further, with $f_2$ reaching a zero-field resonance frequency of about \SI{28}{\giga\hertz} as shown in Fig.~\ref{fig:BBFMR1}(c). The lines in Fig.~\ref{fig:BBFMR1}(b) and (c) are fits to Eq.~\eqref{eq:Kittel} as explained in the following. From the full dataset of VNA-FMR measurements performed on a series of Mn$_\mathrm{2}$Au~(40 nm)/Py~($d_\mathrm{FM}$) bilayers (see SI~\cite{SI}), we extract the zero-field resonance frequencies shown in Fig.~\ref{fig:BBFMR1}(d). Both resonance mode frequencies increase when decreasing $d_\mathrm{FM}$, suggesting an interfacial origin of the resonance frequency enhancement.    
\begin{figure}
\includegraphics{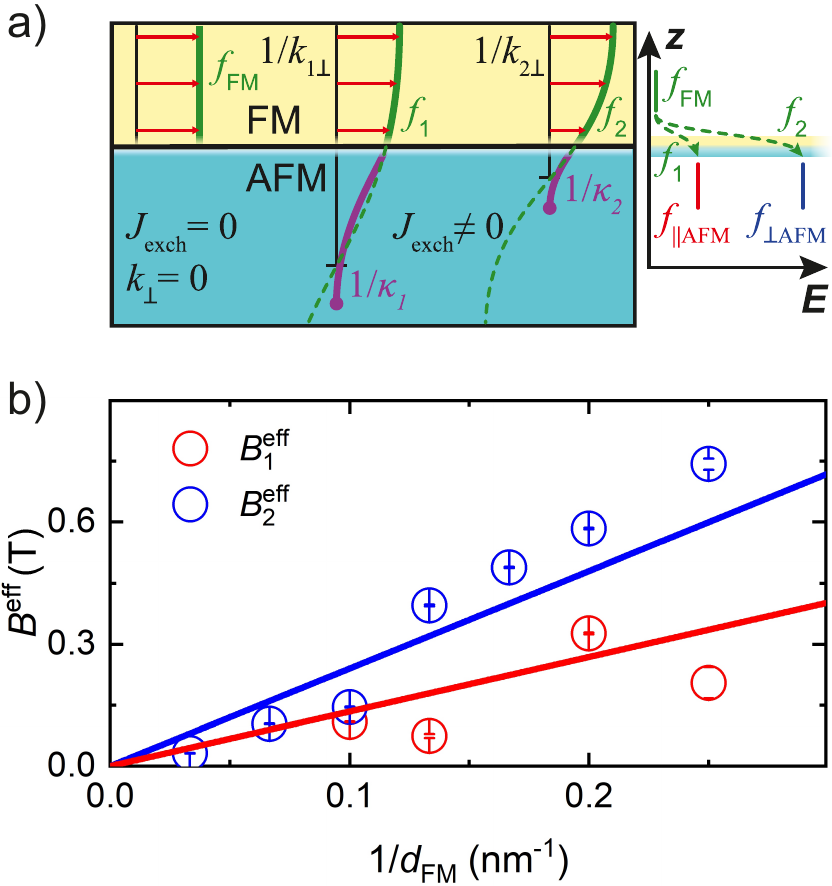}
\caption{(a) Sketch of the sample structure and coupling of modes. The FMR with frequency $f_\mathrm{FM}$ couples to the two antiferromagnetic modes in the AFM. This results in hybrid modes with frequencies $f_1$ and $f_2$ with perpendicular wave-vector $k_{i,\perp}$. The Néel vector excitations decay within $1/\kappa_i$ (see text). (b) The effective stiffness fields resulting from the mode coupling exceed \SI{0.6}{\tesla} for Py thickness $d_\mathrm{FM}\leq\SI{5}{\nano\meter}$.}\label{fig:Banivsd}
\end{figure}

We model the observed dynamics by calculating the magnon spectra of coupled ferro- and antiferromagnetic layers. The essential results of our model are sketched in Fig.~\ref{fig:Banivsd}(a), while a figure depicting the quantitative calculation is shown in the SI~\cite{SI}. The FMR of the ferromagnet becomes evanescent into the AFM layer due to a change of boundary conditions resulting from the coupling of FM and AFM spin dynamics. This results in an effective modification of the FM wavevector $k_\perp$ along the film normal. The derivation of an expression for $k_\perp$ is the main result of our theoretical model calculation explained in the following.  

We start from the standard dipolar-exchange spin-wave dispersion in a tangentially magnetized ferromagnetic thin film given by~\cite{kittel_theory_1948,kalinikos_theory_1986}
\begin{equation}\label{eq:Kittel}
   f(k_\perp,k_\|)=\frac{\gamma}{2\pi} \sqrt{B_\perp B_\parallel}
\end{equation}
with the effective out-of-plane stiffness field
\begin{equation}\label{eq:Bperp}
   B_\perp= \mu_0 H_0 +J_\mathrm{FM}(k_\perp^2+k_\|^2)+\mu_0 M_\mathrm{FM} G_0
\end{equation}
and in-plane stiffness field
\begin{equation}\label{eq:Bparallel}
   B_\parallel=\mu_0 H_0+J_\mathrm{FM}(k_\perp^2+k_\|^2)+\mu_0 M_\mathrm{FM} (1-G_0)\;.
\end{equation}
Here, $M_\mathrm{FM}$ is the saturation magnetization of the ferromagnet. The external magnetic field $\mathbf{H}_0$ is applied in the film plane parallel to the magnetization $\mathbf{M}$. $J_\mathrm{FM}=A_\mathrm{FM} /M_\mathrm{FM}$ with exchange stiffness $A_\mathrm{FM}$. The factor $G_0=(1 - \exp(- k_\| d)) / (k_\| d)$ accounts for the effects of dipole-dipole interactions~\cite{kalinikos_theory_1986}, where $k_\|$ is the in-plane wave vector. In the VNA-FMR experiments discussed above, $k_\|=0$, but $k_\perp\neq0$ as sketched in Fig.~\ref{fig:Banivsd}(a). %Due to confinement conditions, the spatial distribution of the dynamic magnetization corresponds to the PSSW modes in $z$ (out-of-plane) direction with the quantized value of the wave-vector component $k_\perp=\pi n/d$ ($n=0,1,\ldots$). The lowest mode with $n=0$ corresponds to homogeneous magnetization oscillations within the ferromagnetic layer. 

We assume exchange coupling between the magnetization of the ferromagnetic layer, $\mathbf{M}= M_\mathrm{FM}\mathbf{m}$, and the N\'eel vector, $\mathbf{N}=M_\mathrm{AFM}\mathbf{n}$, in a thin layer near the ferro-antiferromagnetic interface. Here $M_\mathrm{FM}$ and $M_\mathrm{AFM}$ are saturation magnetizations of ferro- and antiferromagnetic layer, and both  $\mathbf{m}$ and $\mathbf{n}$ are unit vectors. The corresponding contribution to the energy density of the bilayer due to the interfacial exchange coupling is modelled as $w_\mathrm{int}=-\xi J_\mathrm{exch}M_\mathrm{AFM}M_\mathrm{FM}\mathbf{m}\cdot\mathbf{n}\delta(z)$, where $J_\mathrm{exch}$ is the exchange coupling between FM and AFM, $\xi$ is the thickness of the interfacial region in which the coupling is nonzero, and $\delta(z)$ is the Dirac function. The Py layer of thickness $d_\mathrm{FM}$ is modelled as an easy-plane ferromagnet with negligibly small in-plane anisotropy. Mn$_\mathrm{2}$Au is treated as a two-sublattice easy-plane antiferromagnet with tetragonal magnetic anistropy that sets two equivalent orthogonal easy directions within the film plane. We can assume that due to the strong interfacial exchange coupling $J_\mathrm{exch}$, $\mathbf{M}\|\mathbf{N}$ in equilibrium~\cite{bommanaboyena_readout_2021}. The magnon spectra are calculated based on coupled Landau-Lifshitz equations for ferro-and antiferromagnetic layers (see SI~\cite{SI}). %For ferromagnetic layer we also account for demagnetizing effects, but disregard them for an antiferromagnet. 
The spectra of the antiferromagnetic layer include two magnon branches corresponding to oscillations of the N\'eel vector either in-plane ($f_{\|0}$) or out-of plane ($f_{\perp0}$) with the frequencies  $f_{\|\mathrm{AFM}}=\sqrt{f_{\|0}^2+c^2(k_\perp^2+k_\|^2)}$ and  $f_{\perp\mathrm{AFM}}=\sqrt{f_{\perp0}^2+c^2(k_\perp^2+k_\|^2)}$. Here, $2\pi c$ is the magnon velocity in the antiferromagnet which depends on the AFM exchange stiffness, $A_\mathrm{AFM}$, and the exchange field $B_\mathrm{ex}$ that keeps the antiparallel alignment of magnetic sublattices in the antiferromagnet: $c=(\gamma/2\pi)\sqrt{A_\mathrm{AFM}B_\mathrm{ex}/M_\mathrm{AFM}}$. The hierarchy of frequencies is $f_{\perp0}\gg f_{\|0}\gg f_\mathrm{FM}$, and the effect of the dc magnetic field $\mathbf{H}_0$ on the AFM spectra can be neglected. 
Next, we calculate the magnon spectra assuming nonzero coupling between the ferro- and antiferromagnetic layers. The coupling modifies the boundary conditions at the interface (at $z=0$) as follows:
\begin{eqnarray}\label{eq:BC}%% the sign is related with the direciton of the surface normal!!!!
 && -A_\mathrm{FM}\partial_z  m_{\alpha}+\xi  J_\mathrm{exch}M_\mathrm{AFM}M_\mathrm{FM} (m_{\alpha}-n_\alpha)=0,\nonumber\\ &&A_\mathrm{AFM}\partial_z n _{\alpha}+\xi J_\mathrm{exch}M_\mathrm{AFM}M_\mathrm{FM} (n_\alpha-m_{\alpha})=0,
\end{eqnarray}
where $\alpha$ denotes out-of-plane, $\perp$, or in-plane, $\|$, components of $\mathbf{m}$ and $\mathbf{n}$.
Our calculations show that the eigenmodes consist of the superposition of propagating ferromagnetic magnons and evanescent oscillations of the N\'eel vector in the near-interface region that decay as $\propto \exp\left(-\kappa z\right)$, where $\kappa$ is the decay constant.  Moreover, the spectra also split into two branches, $f_{1}$ and $f_{2}$, corresponding to excitation of either in-plane or out-of-plane oscillations in the antiferromagnetic layer. The decay length $\kappa$ depends on the eigenfrequency of the mode and is different for in-plane and out-of-plane branches: $\kappa_1=\sqrt{f_{\|0}^2-f_{1}^2}/c\approx f_{\| 0}/c$ and $\kappa_2=\sqrt{f_{\perp 0}^2-f_{2}^2}/c\approx f_{\perp 0}/c$. The eigenfrequencies $f_{1}\equiv f(k_\|,k_{1\perp})$ and $f_{2}\equiv f(k_\|,k_{2\perp})$ are calculated from Eq.~\eqref{eq:Kittel} by substituting those values of wave-vectors $k_{1\perp}$ and $k_{2\perp}$ that satisfy the boundary conditions in Eq.~\eqref{eq:BC} (for more details see SI~\cite{SI}). For estimating $k_{1\perp}$ and $k_{2\perp}$, we focus on the lowest modes with $k_{1\perp}, k_{2\perp}\ll \pi/d_\mathrm{FM}$ that have larger overlap with the homogeneous rf magnetic field than the other modes. 
We fit Eq.~\eqref{eq:Kittel} to the experimentally determined resonance frequencies as a function of $H_0$ for each FM layer thickness with $k_{1\,\perp}$ and $k_{2\,\perp}$ as fitting parameters. We determine the FM k-vectors for each resonance frequency branch $f_1$ and $f_2$.
From the boundary conditions in Eq.~\eqref{eq:BC} and Eq.~\eqref{eq:Kittel} we obtain the approximate expression
\begin{equation}\label{eq:kvectors}
   k_{1/2\perp}=\sqrt{\frac{1}{J_\mathrm{FM}}\frac{\xi \kappa_{1,2}J_\mathrm{exch} A_\mathrm{AFM}}{\kappa_{1,2}J_\mathrm{AFM}+\xi J_\mathrm{exch}M_\mathrm{FM}}}\frac{1}{\sqrt{d_\mathrm{FM}}}, 
\end{equation}
where we neglected the difference between $M_\mathrm{FM}$ and $M_\mathrm{AFM}$. The results of accurate numerical calculations of $k_{1/2\perp}$ are provided in the SI~\cite{SI}.
The combinations $B_1^{\mathrm{eff}}\equiv J_\mathrm{FM}k_{1\perp}^2$ and $B_2^{\mathrm{eff}}\equiv J_\mathrm{FM}k_{2\perp}^2$ that appear in Eq.~\eqref{eq:Kittel}  represent the effective stiffness field induced by the coupled ferromagnetic-antiferromagnetic spin dynamics. According to Eq.~\eqref{eq:kvectors}, the stiffness fields scale linearly as $1/d_\mathrm{FM}$, and they are proportional to the exchange coupling $J_\mathrm{exch}$ between 
ferro- and antiferromagnetic layers. We show the stiffness fields of both modes obtained from fitting $f_1$ and $f_2$ vs. $H_0$ by Eq.~\eqref{eq:Kittel} (data points) in Fig.~\ref{fig:Banivsd}(b). The lines are linear fits in accordance with the scaling expected from Eq.~\eqref{eq:kvectors}. 

We note that the two different values of the stiffness fields correspond to the coupling with different antiferromagnetic modes and have a dynamic origin. These values can be related with the effective anisotropy field estimated in~\cite{bommanaboyena_readout_2021} from coercivity measurements. Because the coercive field $\mu_0H_\mathrm{c}$ of our sample series agrees within experimental uncertainty to $B_1^\mathrm{eff}$ (see SI~\cite{SI}) we can estimate
\begin{equation}\label{eq:coercivity}
    B_1^\mathrm{eff}\approx\mu_0 H_\mathrm{c} = \frac{4\xi  B_\mathrm{an} J_{\mathrm{exch}} M_{\mathrm{FM}} d_{\mathrm{AFM}}}{4 B_\mathrm{an} d_{\mathrm{AFM}} +\xi J_{\mathrm{exch}} M_{\mathrm{FM}} } \frac{1}{d_{\mathrm{FM}}},
\end{equation}
where $B_\mathrm{an}$ is the AFM in-plane anisotropy field and $d_\mathrm{AFM}$ is the AFM layer thickness. This expression corresponds to Eq.~(2) of  Ref.~\cite{bommanaboyena_readout_2021}. The stiffness field $B_\mathrm{2}^{\mathrm{eff}}$ is related to the out-of-plane anisotropy of Mn$_2$Au in a similar way. However, this field can be observed only in the magnetic dynamics explored here and not in the static measurements previously performed in~\cite{bommanaboyena_readout_2021}.

Using Eq.~\eqref{eq:coercivity} for fitting the experimentally determined $\mu_0 H_\mathrm{c}$~\cite{SI, bommanaboyena_readout_2021}, we determine $\xi J_{\mathrm{exch}} M_{\mathrm{FM}}\approx1.6\,\mathrm{T\,nm}$. Using fitting of the effective fields $B_1^{\mathrm{eff}}\equiv J_\mathrm{FM}k_{1\perp}^2$ and $B_2^{\mathrm{eff}}\equiv J_\mathrm{FM}k_{2\perp}^2$ as a function of the inverse FM thickness, as shown in Fig.~\ref{fig:Banivsd}(b), we further determine $B_2^{\mathrm{eff}} / B_1^{\mathrm{eff}} \approx 2.4$. This ratio is inserted into Eq.~\eqref{eq:kvectors}, from which we obtain
\begin{equation}\label{eq:fitkvectors}
    \frac{B_2^{\mathrm{eff}}}{B_1^{\mathrm{eff}}}=\frac{\kappa_2}{\kappa_1} \cdot\frac{{ \kappa_1 J_{\mathrm{AFM}}}+\xi J_{\mathrm{exch}} M_{\mathrm{FM}}}{\kappa_2 J_{\mathrm{AFM}} +\xi J_{\mathrm{exch}}M_{\mathrm{FM}} }.
\end{equation}
From Eq.~\eqref{eq:fitkvectors} we can estimate
 \begin{equation}\label{eq:kappa}
  \frac{1}{\kappa_1}- \frac{3}{\kappa_2} =\left( \frac{B_2^\mathrm{eff}}{B_1^\mathrm{eff}}-1\right)\frac{J_\mathrm{AFM}}{\xi J_\mathrm{exch}M_\mathrm{FM} }\approx 11 \,\mathrm{nm}.
\end{equation}
Taking into account that $\kappa_1\ll \kappa_2$ and using the value of magnon velocity 
%$2\pi c=20$ m/s 
$2\pi c=\SI{22.49}{\kilo\meter\per\second}$ (see SI~\cite{SI}), we estimate $f_{\| 0}\approx \SI{0.3}{\tera\hertz}$, which is about a factor two larger than previous direct THz measurements of magnons in Mn$_2$Au films grown by a different technique~\cite{Arana2017}. 

%where we estimated $\kappa_1$ and $\kappa_2$ as mentioned earlier. From Eqn.~\eqref{eq:fitkvectors} we 
%$\mathcal{O}(1~\mathrm{T})$ with $f_{\perp 0}\approx \SI{1}{\tera\hertz}$ and $f_{\| 0}\approx \SI{0.1}{\tera\hertz}$.

%we first note that the stiffness field of the in-plane mode can be expressed in terms of the domain wall width $x_\mathrm{DW}$ and in-plane magnetic anisotropy field $H_\mathrm{AFM}$ of the Mn$_2$Au layer as $B_\mathrm{2}^{\mathrm{eff}}\propto H_\mathrm{AFM}x_\mathrm{DW}/d_\mathrm{FM}$. $x_\mathrm{DW}$ is the penetration depth of the magnetic correlations between ferromagnetic magnetization and the N\'eel vector. If the thickness of Mn$_2$Au layer is smaller or of the order of $x_\mathrm{DW}$, 

\begin{figure}
\centering
\includegraphics{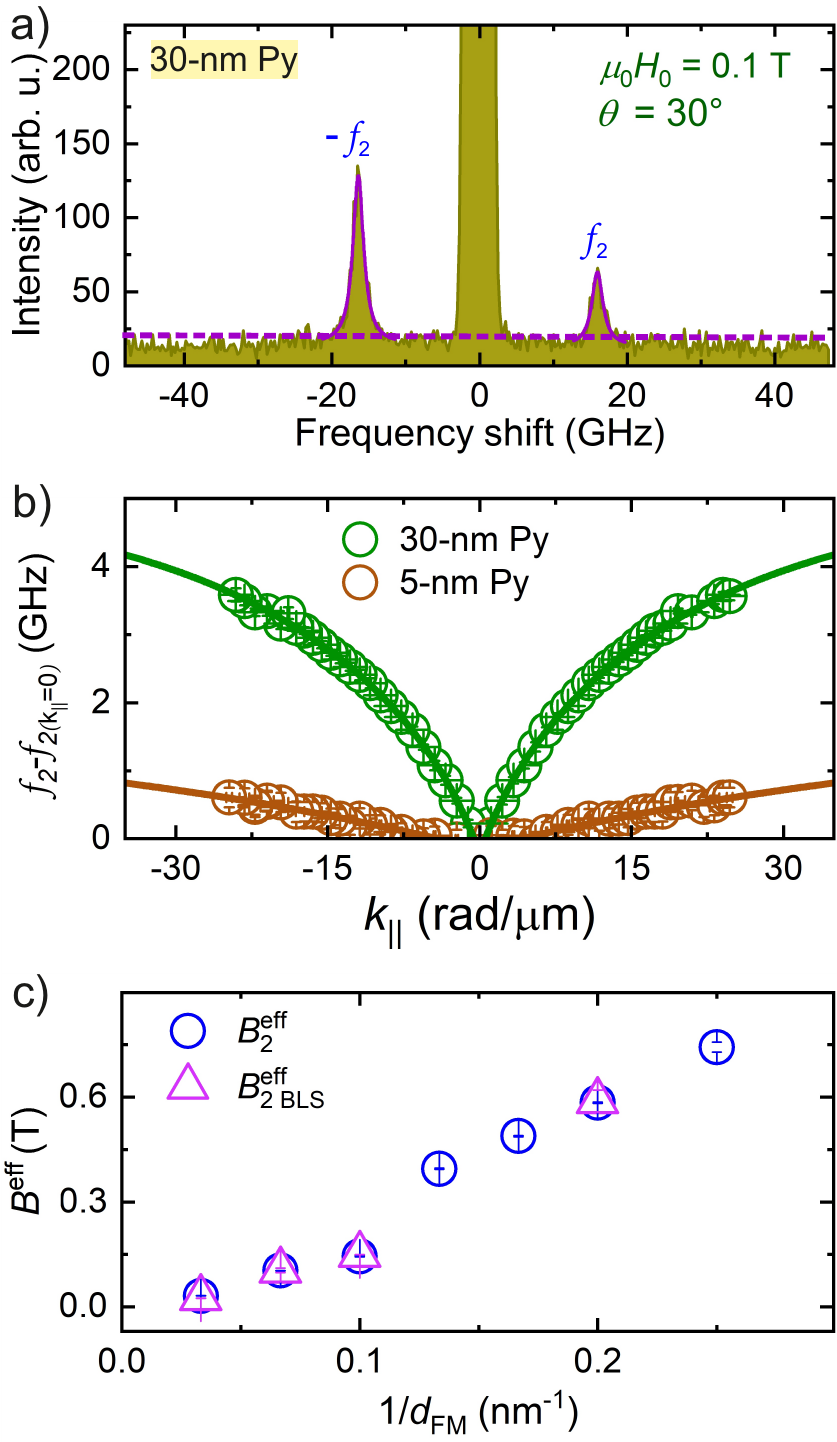}
\caption{ (a) BLS spectrum. The solid lines are Lorentzian fits used to determine $f_2$, the dashed line represents the background signal. (b) The spin-wave dispersion of samples with \SI{5}{\nano\meter} and \SI{30}{\nano\meter} Py obtained by fitting the BLS spectra (points) and fit to Eq.~\eqref{eq:Kittel} (lines). Spectra are shifted on the y-axis by the  $f_2(k_\parallel=0)$ mode frequency for clarity. (c) The stiffness field $B_2^{\mathrm{eff}}$ extracted from fitting the BLS measurements is in full agreement with that observed in FMR measurements.}\label{fig:BLS}
\end{figure}

To separate the impact of $k_\parallel$ and $k_\perp$ on the hybrid mode dynamics, we carried out additional wave-vector resolved BLS measurements of thermal magnons~\cite{sebastian_micro-focused_2015, sandercock_light_1978,DEMOKRITOV2001441,ordonez-romero_three-magnon_2009, kieffer_brillouin_2015}. 
 
Fig.~\ref{fig:BLS}(a) shows an exemplary BLS spectrum. The peaks at $\pm f_2$ in the BLS spectra are fitted by Lorentzian functions to obtain the dependence of $f_2$ on in-plane wave-vector shown  in Fig.~\ref{fig:BLS}(b) for samples with \SI{5}{\nano\meter} and \SI{30}{\nano\meter} thick Py layers. The signal-to-noise ratio only allowed us to measure the $f_2$ mode by BLS. We fit the spin-wave dispersions to Eq.~\eqref{eq:Kittel} and obtain the stiffness fields $B_2^{\mathrm{eff}}$ shown in Fig.~\ref{fig:BLS}(c) that are in excellent agreement with our FMR results. The agreement of the stiffness fields observed in BLS and FMR demonstrates that the coupling of AFM and FM modes impacts $k_\perp$ and not $k_\parallel$ in accordance with the assumptions in our model.

In summary, we have made a detailed investigation of the magnetization dynamics in exchange-coupled bilayers of Mn$_\mathrm{2}$Au/Py. We demonstrate that the interfacial exchange coupling in the ferro/antiferromagnetic bilayer system enables the control of the hybrid-mode resonance frequency and spin-wave dispersion by variation of the thickness of the ferromagnetic layer. The splitting of the unperturbed ferromagnetic resonance frequency of Py into two non-degenerate modes with strongly enhanced frequency can be understood in the context of coupling of Py dynamics to the in-plane and out-of-plane modes of the easy-plane antiferromagnet. The same modes are observed also in the thermal magnon spectrum recorded by BLS spectroscopy, demonstrating that the coupling is independent of the in-plane $k$-vector of the spin waves in the hybrid system. The magnitude of the frequency enhancement depends on the strength of the interlayer exchange coupling and the magnon frequencies of the AFM. By independent determination of the interlayer exchange coupling strength, our method allows to estimate the AFM magnon frequencies even though they lie far above the regime experimentally accessible to us. Exploiting THz AFM dynamics to control sub-THz hybrid spin dynamics may find applications in high-frequency devices such as spin-torque oscillators.

\begin{acknowledgments}
We acknowledge financial support by the Deutsche Forschungsgemeinschaft (DFG, German Research Foundation) within the Transregional Collaborative Research Center TRR 173/2–268565370 “Spin+X” (Projects B13, A01, A03, A05, A11, B02, B11, and B12). M.J. and M. K. acknowledge support by EU HORIZON-CL4-2021-DIGITAL-EMERGING-01-14 programme under grant agreement No. 101070287 (SWAN-on-Chip).  O.G. and T.W. also acknowledge support from Deutsche Forschungsgemeinschaft (DFG) SPP 2137, project number 403233384.  J.S. additionally acknowledges funding from the Grant Agency of the Czech Republic grant no. 19-28375X. M.K. acknowledges support from the Horizon 2020 and Horizon Europe Framework of the European Commission under Grant No.863155 (S-NEBULA). M.W. acknowledges support from the European Research Council (ERC) under the European Union’s Horizon Europe research and innovation programme (Grant agreement No. 101044526). 
\end{acknowledgments}

\bibliographystyle{apsrev4-2}
\bibliography{Manuscript_Mn2Au-Py}% Produces the bibliography via BibTeX.

\end{document}